\documentclass[runningheads]{llncs}
\usepackage[T1]{fontenc}
\usepackage{booktabs}
\usepackage{graphicx}
\usepackage{bibunits}
\usepackage{multirow}
\usepackage[misc]{ifsym}
\usepackage{booktabs}
\usepackage{amsmath,amssymb,amsfonts}
\usepackage[dvipsnames]{xcolor}
\usepackage{bm}

\usepackage[colorlinks=true,linkcolor=blue,citecolor=blue,urlcolor=blue]{hyperref}
\usepackage{color}

\definecolor{mygreen}{RGB}{87, 176, 98}
\definecolor{myred}{RGB}{235, 70, 56}
\definecolor{myyellow}{RGB}{245, 198, 79}
\begin{document}
\title{
Artifact Restoration in Histology Images with Diffusion Probabilistic Models}
\author{
Zhenqi He\inst{1}\orcidID{0009-0000-2265-7159}\and Junjun He\inst{2}\and Jin Ye\inst{2}\and Yiqing Shen\inst{3}\textsuperscript{(\Letter)}\orcidID{0000-0001-7866-3339}
}
%index{He, Zhenqi}
%index{He, Junjun}
%index{Ye, Jin}
%index{Shen, Yiqing} 
\authorrunning{Z.He et al.}
\institute{
\textsuperscript{1}The University of Hong Kong\\
\textsuperscript{2}Shanghai AI Laboratory\\
\textsuperscript{3}Johns Hopkins University\\
\email{yshen92@jhu.edu}
}
\maketitle              

\begin{abstract}
Histological whole slide images (WSIs) can be usually compromised by artifacts, such as tissue folding and bubbles, which will increase the examination difficulty for both pathologists and Computer-Aided Diagnosis (CAD) systems. 
Existing approaches to restoring artifact images are confined to Generative Adversarial Networks (GANs), where the restoration process is formulated as an image-to-image transfer.
Those methods are prone to suffer from mode collapse and unexpected mistransfer in the stain style, leading to unsatisfied and unrealistic restored images.
Innovatively, we make the first attempt at a denoising diffusion probabilistic model for histological artifact restoration, namely \texttt{ArtiFusion}.
Specifically, \texttt{ArtiFusion} formulates the artifact region restoration as a gradual denoising process, and its training relies solely on artifact-free images to simplify the training complexity.
Furthermore, to capture local-global correlations in the regional artifact restoration, a novel Swin-Transformer denoising architecture is designed, along with a time token scheme. 
Our extensive evaluations demonstrate the effectiveness of \texttt{ArtiFusion} as a pre-processing method for histology analysis, which can successfully preserve the tissue structures and stain style in artifact-free regions during the restoration.
Code is available at \url{https://github.com/zhenqi-he/ArtiFusion}.
\end{abstract}

\section{Introduction}
Histology is critical for accurately diagnosing all cancers in modern medical imaging analysis. 
However, the complex scanning procedure for histological whole-slide images (WSIs) digitization may result in the alteration of tissue structures, due to improper removal, fixation, tissue processing, embedding, and storage~\cite{artifacts}. 
Typically, these changes in tissue details can be caused by various extraneous factors such as bubbles, tissue folds, uneven illumination, pen marks, altered staining, and \textit{etc}~\cite{artifacttype}.
Formally, the changes in tissue structures are known as artifacts.
The presence of artifacts not only makes the analysis more challenging for pathologists but also increases the risk of misdiagnosis for Computer-Aided Diagnosis (CAD) systems~\cite{artifactresult}. 
Particularly, deep learning models, which have become increasingly prevalent in histology analysis, have shown vulnerability to the artifact, resulting in a two-times increase in diagnosis errors~\cite{zhang2022benchmarking}.

\begin{figure}[t!]
\centering
\includegraphics[width=0.85\textwidth]{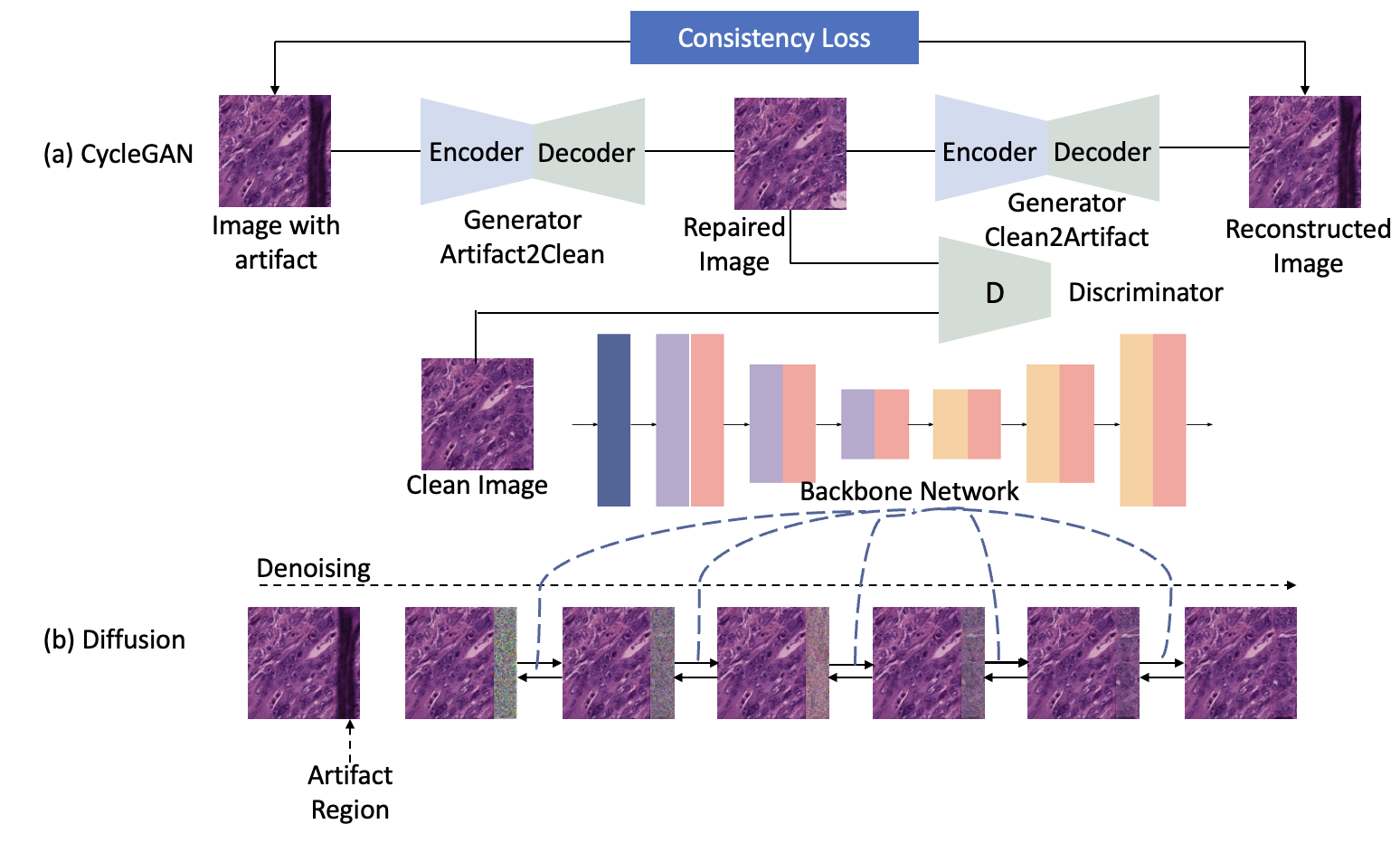}
\caption{ 
Learning-based artifact restoration approaches.
(a) CycleGAN~\cite{cycleGAN} formulates the artifact restoration as an image-to-image transfer problem.
It leverages two pairs of the generator and discriminator to learn the transfer between the artifact and artifact-free image domains.
(b) Diffusion probabilistic model~\cite{DDPM} (ours) formulates artifact restoration as a regional denoising process.
}
\label{fig:intro}
\end{figure}

In real clinical practice, rescanning the WSIs that contain artifacts can partially address this issue.
However, it may require multiple attempts before obtaining a satisfactory WSI, which can lead to a waste of time, medical resources, and deplete tissue samples.
Discarding the local region with artifacts for deep learning models is another solution, but it may result in the loss of critical contextual information.
Therefore, learning-based artifact restoration approaches have gained increasing attention.
For example, CycleGAN~\cite{cycleGAN} formulates the artifact restoration as an image-to-image transfer problem by learning the transfer between the artifact and artifact-free image domains from unpaired images, as depicted in Fig.~\ref{fig:intro}(a).
However, existing artifact restoration solutions are confined to Generative Adversarial Networks (GANs)~\cite{GANs}, which are difficult to train due to the mode collapse and are prone to suffer from unexpected stain style mistransfer.
To address these issues, we make the first attempt at a diffusion probabilistic model for artifact restoration approach~\cite{DDPM}, as shown in Fig.~\ref{fig:intro}(b).  
Innovatively, our framework formulates the artifact restoration as a regional denoising process, which thus can to the most extent preserve the stain style and avoid the loss of contextual information in the non-artifact region.
Furthermore, our approach is trained solely with artifact-free images, which reduces the difficulty in data collection.

The major contributions are two-fold.
(1) 
We make the first attempt at a denoising diffusion probabilistic model for artifact removal, called \texttt{ArtiFusion}. 
This approach differs from GAN-based methods that require either paired or unpaired artifacts and artifact-free images, as our \texttt{ArtiFusion} relies solely on artifact-free images, resulting in a simplified training process.
(2) To capture the local-global correlations in the gradual regional artifact restoration process, we innovatively propose a Swin-Transformer denoising architecture to replace the commonly-used U-Net and a time token scheme for optimal Swin-Transformer denoising. 
Extensive evaluations on real-world histology datasets and downstream tasks demonstrate the superiority of our framework in artifact removal performance, which can generate reliable restored images while preserving the stain style.  

\section{Methodology}

\begin{figure}[t!]
\centering
\includegraphics[width=0.8\textwidth]{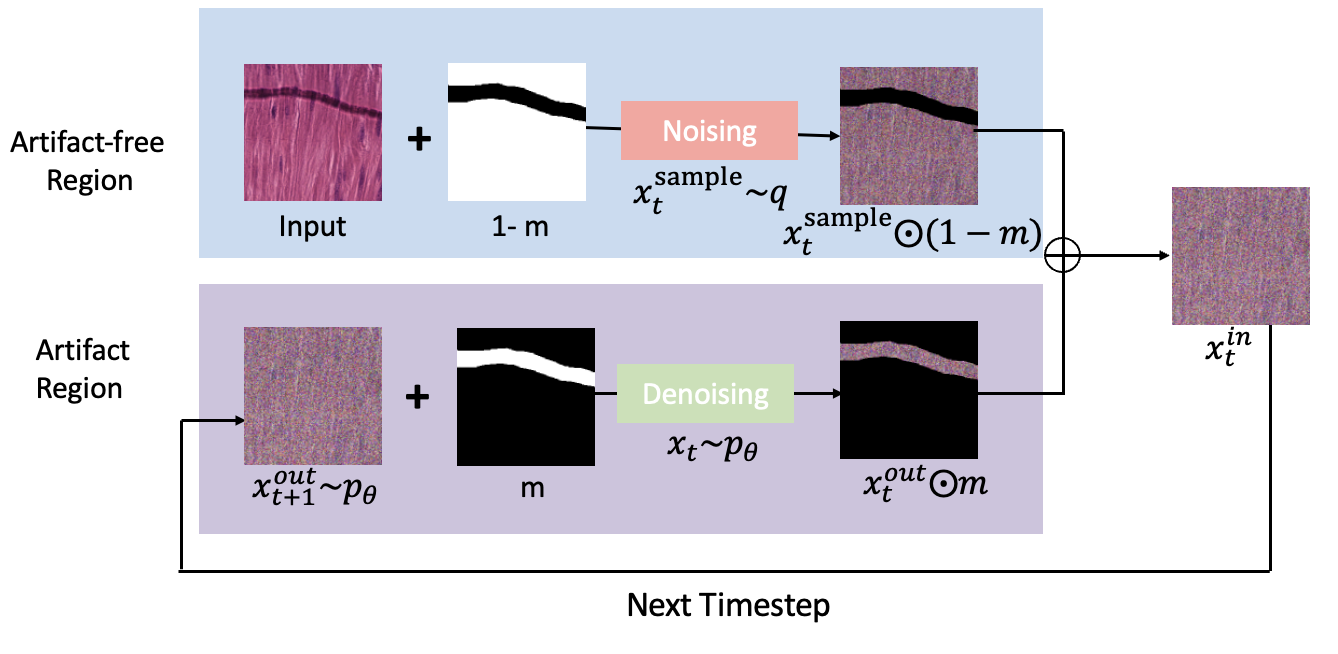}
\caption{The semantic illustration of inference stage in \texttt{ArtiFusion} for local regional artifact restoration.}
\label{fig:inf}
\end{figure}

\subsubsection{Overall Pipeline.}
The proposed histology artifact restoration diffusion model \texttt{ArtiFusion}, comprises two stages, namely the training, and inference. 
During the training stage, \texttt{ArtiFusion} learns to generate regional histology tissue structures based on the contextual information from artifact-free images.
In the inference stage, \texttt{ArtiFusion} formulates the artifact restoration as a gradual denoising process.
Specifically, it first replaces the artifact regions with Gaussian noise, and then gradually restores them to artifact-free images using the contextual information from nearby regions.

\subsubsection{Diffusion Training Stage.}
The proposed \texttt{ArtiFusion} learns the capability of generating local tissue representation from contextual information during the training stage.
To achieve this, we follow the formulations of DDPM~\cite{DDPM}, which involve a forward process that gradually injects Gaussian noise into an artifact-free image and a reverse process that aims to reconstruct images from noise. 
During the forward process, we can obtain a noisy version of $\mathbf{x}_t$ for arbitrary timestep $t \in \mathbb{N}[0,T]$ using a Gaussian transition kernel $q(\mathbf{x}t|\mathbf{x}{t-1}) = \mathcal{N}(x_t;\sqrt{1-\beta_t}\mathbf{x}{t-1},\beta_t\mathbf{I})$, where $\beta_t \in (0,1)$ are predefined hyper-parameters~\cite{DDPM}. 
Simultaneously, the reverse process trains a denoising network $p_{\theta}(\mathbf{x}_{t-1}|\mathbf{x}_{t}^{in})$, which is parameterized by $\theta$, to reverse the forward process $q(\mathbf{x}_{t}|\mathbf{x}_{t-1})$.
The overall training objective $L$ is defined as the variational lower bound of the negative log-likelihood, given by:
\begin{equation}
    \mathbb{E}[-\log p_{\theta}(\mathbf{x}_{0})] 
    \leq 
    \mathbb{E}_{q}[-\log p(\mathbf{x}_{T})-\sum_{1\leq t \leq T}\log\frac{p_{\theta}(\mathbf{x}_{t-1}|\mathbf{x}_{t})}{q(\mathbf{x}_{t}|\mathbf{x}_{t-1})}] =L.
\end{equation}
This formulation is extended in DDPM \cite{DDPM} to be further written as:
\begin{equation}
\small
\nonumber
    L=
    \mathbb{E}_{q}
    [\underbrace{D_{KL}(q(\mathbf{x}_{T}|x_{0}))||p(\mathbf{x}_{T})}_{L_{T}} 
    + \sum_{t>1}
    \underbrace{D_{KL}(q(\mathbf{x}_{t-1}|\mathbf{x}_{t},\mathbf{x}_{0}))||p_{\theta}(\mathbf{x}_{t-1}|\mathbf{x}_{t})}_{L_{t-1}}-\underbrace{\log p_{\theta}(\mathbf{x}_{0}|\mathbf{x}_{1})}_{L_{0}}],
\end{equation}
where $D_{KL}(\cdot||\cdot)$ is the KL divergence.

\begin{figure}[t!]
\centering
\includegraphics[width=0.93\textwidth]{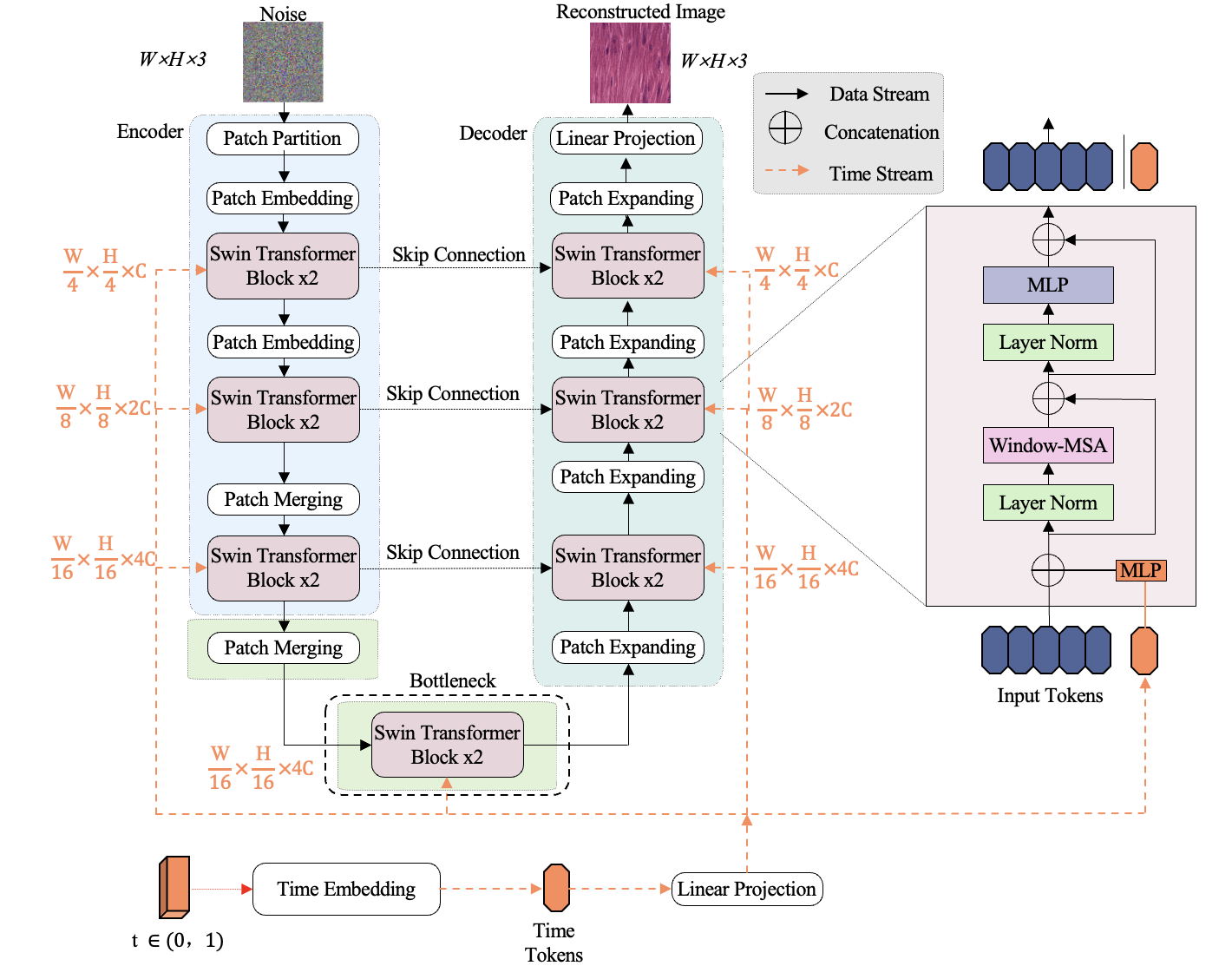}
\caption{The proposed Swin-Transformer denoising network. 
}
\label{fig:model}
\end{figure}

\subsubsection{Artifact Restoration in Inference Stage.}
During the inference stage, we first use a threshold method to detect the artifact region in the input image $\mathbf{x}_0$.
Then, unlike the conventional diffusion models~\cite{DDPM} that aim to generate the entire image, \texttt{ArtiFusion} selectively performs denoising resampling only in the artifact region to maximally preserve the original morphology and stain style in the artifact-free region, as shown in Fig.~\ref{fig:inf}.
Specifically, we represent the artifact-free region and the artifact region in the input image as $\mathbf{x}_0 \odot (1-\mathbf{m})$ and $\mathbf{x}_0 \odot \mathbf{m}$, respectively~\cite{repaint}, where $\mathbf{m}$ is a Boolean mask indicating the artifact region and $\odot$ is the pixel-wise multiplication operator.
To perform the denoising resampling, we write the input image $ \mathbf{x}_{t}^{in}$ at each reverse step from $t$ to $t-1$ as the sum of the diffused artifact-free region and the denoised artifact region, \textit{i.e.,}
\begin{equation}
    \mathbf{x}_{t}^{in} = 
    \mathbf{x}_{t}^{sample} \odot (1-\mathbf{m})
    +
   \mathbf{x}_{t+1}^{out} \odot \mathbf{m}, 
    \label{eq:reverse}
\end{equation}
where $\mathbf{x}_{t}^{sample}o\odot (1-\mathbf{m})$ is artifact-free region diffused for $t$ times using the Gaussian transition kernel \textit{i.e.} $\mathbf{x}_{t}^{sample}\sim \mathcal{N}(\sqrt{\bar{\alpha}_t}\mathbf{x}_0, (1-\bar{\alpha}_t \mathbf{I}))$ with $\bar{\alpha}_t=\prod_{i=1}^t(1-\beta_i)$; 
and $\mathbf{x}_{t+1}^{out}$ is the output from the denoising network in the previous reverse step \textit{i.e.,} $p_{\theta}(\mathbf{x}_{t+1}^{out}|\mathbf{x}_{t+1}^{in})$.
Consequently, the final restored image is obtained as $\mathbf{x}_{0} \odot (1-\mathbf{m})+\mathbf{x}_{0}^{out} \odot \mathbf{m}$.

\subsubsection{Swin-Transformer Denoising Network.}
To capture the local-global correlation and enable the denoising network to effectively restore the artifact regions, we propose a novel Swin-Transformer-basedr~\cite{SwinTransformer} denoising network for \texttt{ArtiFusion}.
As shown in Fig.~\ref{fig:model}, our network follows a U-shape architecture, where the encoder, bottleneck, and decoder modules all employ Swin-Transformer as the basic building block.
Additionally, we introduce an innovative auxiliary time token to inject the time information.
In an arbitrary time step $t$ during the training process, to obtain a time token, we first embed the scalar $t$ by learnable linear layers, with weights that are specific to each Swin-Transformer block.
In contrast to existing U-Net based denoising networks~\cite{DDPM}, we propose a better interaction between hidden features and time information by concatenating the time token to feature tokens before passing them to the attention layers. 
The resulting tokens are then processed by the attention layers, and the auxiliary time token is discarded to retain the original feature dimension to fit the Swin-Transformer block design after the attention layers.

\section{Experiments}
\subsubsection{Dataset.}
To evaluate the performance of artifact restoration, a training set is curated from a subset of Camelyon17~\cite{litjens20181399} \footnote{Available at \url{https://camelyon17.grand-challenge.org}.}. 
It comprises a total number of 2445 artifact-free images and another 2547 images with artifacts, where all histological images are scaled to the resolution of $256 \times 256$ pixels at the magnitude of $20\times$. 
The test set uses another public histology image dataset~\cite{ClusterSeg} with 462 artifact-free images\footnote{Available at \url{https://github.com/lu-yizhou/ClusterSeg}.}, where we obtain the paired artifact images by the manually-synthesized artifacts~\cite{zhang2022benchmarking}.

\begin{figure}[t!]
\centering
\includegraphics[width=1\textwidth]{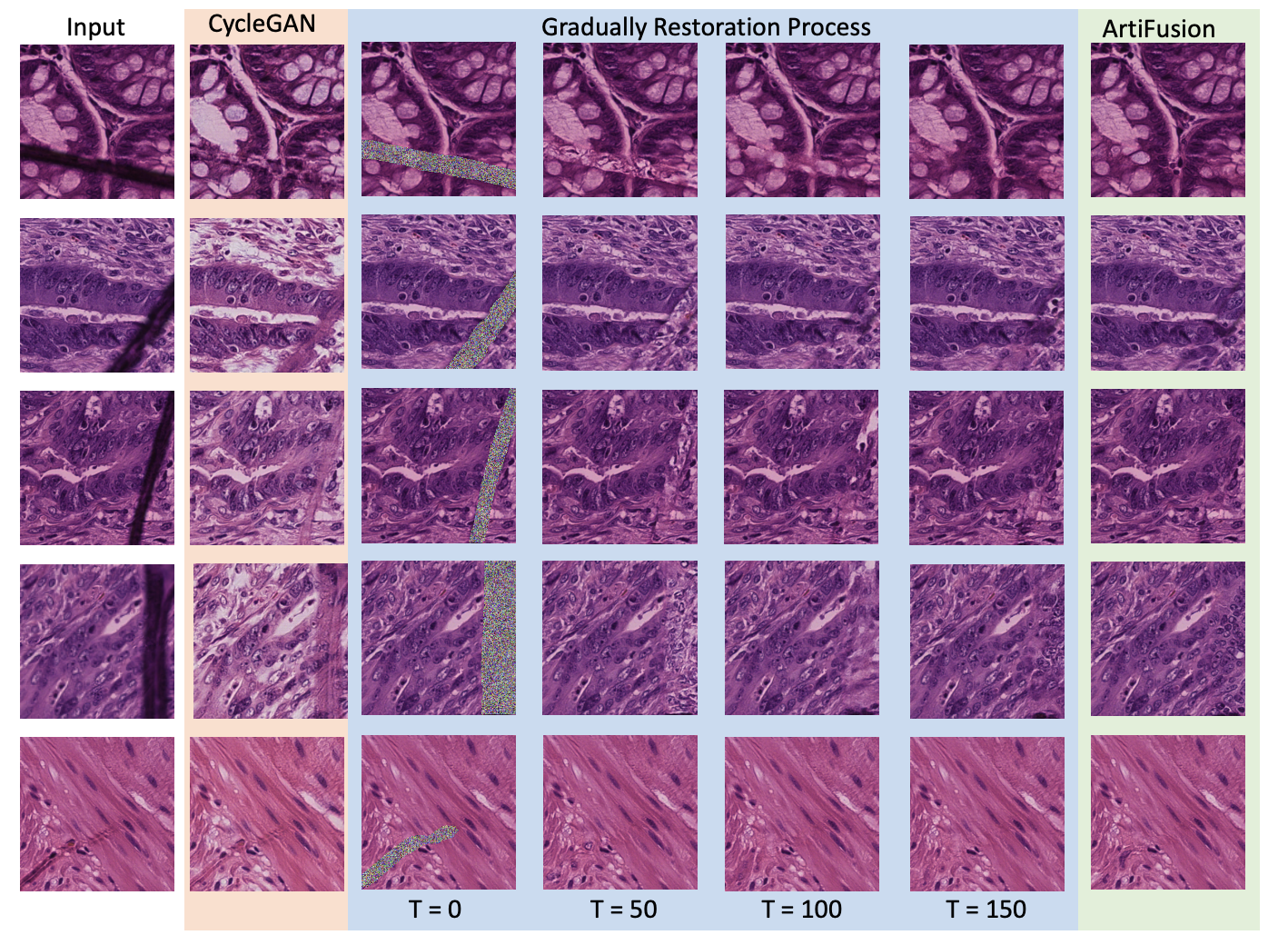}
\caption{ 
Artifact restoration on five real-world artifact images.
We observe that \texttt{ArtiFusion} can successfully overcome the drawback of stain style mistransfer in CycleGAN.
We also illustrate the gradual denoising process in the artifact region by \texttt{ArtiFusion}, at time step $t=0,50,100,150$.
It highlights the ability of \texttt{ArtiFusion} to progressively remove artifacts from the histology image, resulting in a final restored image that is both visually pleasing and scientifically accurate.
}
\label{fig:result}
\end{figure}

\subsubsection{Implementations.}
We implement the proposed \texttt{ArtiFusion} and its counterpart in Python 3.8.10 and PyTorch 1.10.0.
All experiments are carried out in parallel on two NVIDIA RTX A4000 GPU cards with 16 GiB memory. 
Hyper-parameters are as follows: a learning rate of $10^{-4}$ with Adam optimizer, the total timesteps is set to $250$.

\subsubsection{Compared Methods and Evaluation Metrics.}
As a proof-of-concept attempt at a generative-models-based artifact restoration framework in the histology domain, currently, there are limited available literature works and open-sourced codes for comparison. 
Consequently, we leverage the prevalent CycleGAN~\cite{cycleGAN} as the baseline for comparison, because of its excellent performance in the image transfer, and also its nature that requires no paired data can fit our circumstance. 
Unlike CycleGAN which requires both artifact-free images and artifact images, \texttt{ArtiFusion} only relies on artifact-free images, leading to a size of the training set that is half that of CycleGAN.
For a fair compaison, we train the CycleGAN with two configurations, namely (\#1) using the entire dataset, and (\#2) using only half the dataset, where the latter uses the same number of the training samples as \texttt{ArtiFusion}.
Regarding the ablation, we compare the proposed Swin-Transformer denoising network with the conventional U-Net~\cite{DDPM} (denoted as `U-Net'), and the time token scheme with the direct summation scheme (denoted as `Add').
We use the following metrics: $L_2$ distance (L2) with respect to the artifact region, the mean-squared error (MSE) over the whole image, structural similarity index (SSIM)~\cite{SSIM}, Peak signal-to-noise ratio (PSNR)~\cite{PSNR}, Feature-based similarity index (FSIM)~\cite{zhang2011fsim} and Signal to reconstruction error ratio (SRE)~\cite{SRE}.

\begin{table}[t!]
\centering
\caption{
Quantitative comparison of \texttt{ArtiFusion} with CycleGAN on artifact restoration performance. 
The $\downarrow$ indicates the smaller value, the better performance; and vice versa. 
}
\label{tab:result}
\begin{tabular}{l|c|c|c|c|c|c}
% \hline
\toprule
Methods & L2~($\times10^4$) $\downarrow$ & MSE~$\downarrow$ & SSIM~$\uparrow$ & PSNR~$\uparrow$ & FSIM~$\uparrow$ & SRE~$\uparrow$ \\ 
\hline
CycleGAN (\#1)~\cite{cycleGAN} & $1.119$ & $0.5583$ &  $0.9656$     &  $42.37$   & $0.7188$ & $51.42$    \\
CycleGAN (\#2)~\cite{cycleGAN} & $1.893$ & $0.5936$ &  $0.9622$     &  $42.12$  & $0.7162$ & $50.21$     \\
\hline
\texttt{ArtiFusion} (U-Net) & $0.5027$ & $0.2508$ &  $0.9850$     &    $47.61$   & $0.8173$ &  $54.59$  \\
\texttt{ArtiFusion} (Add) &  $0.5007$   &  $0.2499$ &  $0.9850$    &   $47.79$  & $0.8184$ & $54.76$   \\ 
\texttt{ArtiFusion} (Full Settings)    &   \bm{$0.4940$}   & \bm{$0.2465$}  &  \bm{$0.9860$}    & \bm{$48.08$} & \bm{$0.8216$} & \bm{$55.43$}       \\ 

% \hline
\bottomrule
\end{tabular}
\end{table}

\begin{table}[b!]
\centering
\caption{
Comparison of the model complexity and efficiency in terms of the number of parameters, FLOPs, and averaged inference time.
}
\label{tab:cost}
\begin{tabular}{l|c|c|c}
% \hline
\toprule
Methods & \#Params~($\times10^6$) & FLOPs~($\times10^9$) & Inference(s) \\ 
\hline
CycleGAN \cite{cycleGAN} & $28.28$ & $60.04$ & $1.065$ \\
\hline
\texttt{ArtiFusion} (UNet) & $108.41$ & $247.01$  & $112.37$ \\ 
\texttt{ArtiFusion} (Add)  & $27.74$ & $7.69$ & $30.14$\\ 
\texttt{ArtiFusion} (Full Settings) & $29.67$ & $7.73$ &  $30.71$\\ 

% \hline
\bottomrule
\end{tabular}
\end{table}

\subsubsection{Evaluations on Artifact Restoration.} 
The quantitative comparison with CycleGAN and \texttt{ArtiFusion} are shown in Table~\ref{tab:result}, where some exemplary images are illustrated in Fig.~\ref{fig:result}. 
Our results demonstrate the superiority of \texttt{ArtiFusion} over GAN in the context of artifact restoration, with a large margin observed in all evaluation metrics. 
For instance, \texttt{ArtiFusion} can reduce the L2 and MSE by more than $50\%$, namely from $1\times 10 ^4$ to $0.5\times10 ^4$ and from $0.55$ to $0.25$ respectively.
It implying that our method can to the large extent restore the artifact regions using the global information.
In addition, \texttt{ArtiFusion}  can improve other metrics, including SSIM, PSNR, FSIM and SRE by $0.0204$, $5.72$, $0.1028$ and $4.02$ respectively,  indicating that it can preserve the stain style during the restoration process. 
Moreover, our ablation study shows that the Swin-Transformer denoising network can outperform the conventional U-Net, highlighting the significance of capturing global correlation for local artifact restoration.
Finally, the concatenating time token with feature tokens can bring an improvement in terms of all evaluation matrices, making it a better fit for the transformer architecture than the direct summation scheme in U-Net~\cite{DDPM}.
In summary, our ablations confirm the effectiveness of all the components in our method.

\begin{table}[t!]
\centering
\caption{
The effectiveness of the proposed artifact restoration framework in the downstream task-tissue classification task.
We report the classification accuracy on the test set (\%) with different network architectures including ResNet \cite{ResNet}, RexNet \cite{rexnet} and EfficientNet \cite{tan2020efficientnet}.
}
\label{tab:classification}
\resizebox{\linewidth}{!}{ 
\begin{tabular}{l|c|c|c|c|c}
% \hline
\toprule
Settings &  ResNet18   & ResNet34  & ResNet50 & RexNet100  & EfficientNetB0   \\ 
\hline
Clean  & $95.529$ & $93.538$ &    $94.833$  & $95.487$   &  $95.808$   \\
\hline
Artifacts & $80.302$ & $86.031$ &  $85.012$     &  $90.446$  & $90.626$  \\
\hline
Restored w CycleGAN  & $86.326$ & $88.273$& $87.994$ & $90.776$  & $91.811$\\
\hline
Restored w \texttt{ArtiFusion}  & $92.376$ &  $91.252 $&  $90.408$   &  $92.310$    &  $94.232$ \\
\bottomrule
\end{tabular}
}
\end{table}

\subsubsection{Comparisons of Model Complexity.}
In Table~\ref{tab:cost}, we compare the model complexity in terms of the number of parameters, Floating Point Operations Per second (FLOPs), and averaged inference time on one image. 
Our proposed model achieves a significant reduction in the number of parameters by $72.6\%$, namely from $108.41 \times 10^6$ to $29.67 \times 10^6$, compared with CycleGAN.
This reduction in model size comes at the cost of longer inference time. 
However, a smaller model size can facilitate easier deployment in real clinical practice.

\subsubsection{Evaluations by Downstream Classification Task.}
We further evaluate the proposed artifact restoration framework on a downstream tissue classification task. 
To this end, we use the public dataset NCT-CRC-HE-100K for training and CRC-VAL-HE-7K for testing, which together contains $100,000$ training samples and $7,180$ test samples. 
We consider the performance on the original unprocessed data, denoted as `Clean', as the upper bound.
Then, we manually synthesize the artifact (denoted as `Artifact') and evaluate the classification performance with restoration approaches CycleGAN and \texttt{ArtiFusion}.
In Table~\ref{tab:classification}, comparisons show that the presence of artifacts can result in a significant performance decline of over 5\% across all five network architectures. 
Importantly, the classification accuracy on images restored with \texttt{ArtiFusion} is consistently higher than those restored with CycleGAN, demonstrating the superiority of our model.
These results highlight the effectiveness of \texttt{ArtiFusion} as a practical pre-processing method for histology analysis.

\section{Conclusion}
In this paper, we propose \texttt{ArtiFusion}, the first attempt at a diffusion-based artifact restoration framework for histology images. 
With a novel Swin-Transformer denoising backbone, \texttt{ArtiFusion} is able to restore regional artifacts using the context information, while preserving the tissue structures in artifact-free regions as well as the stain style. 
Experimental results on a public histological dataset demonstrate the superiority of our proposed method over the state-of-the-art GAN counterpart. 
Consequently, we believe that our proposed method has the potential to benefit the medical community by enabling more accurate diagnosis or treatment planning as a pre-processing method for histology analysis. 
Future work includes investigating the extension of \texttt{ArtiFusion} to more advanced diffusion models such as score-based or score-SDE models~\cite{yang2022diffusion}.

\bibliographystyle{plain}
\bibliography{paper943/paper943_6_ref}

\end{document}